\documentclass[reprint,pra,amsmath,amssymb,floatfix,aps]{revtex4-2}
\usepackage{graphicx}
\usepackage{bm}
\usepackage{color}
\usepackage{amsmath}
\usepackage{amssymb}

\begin{document}

\author{Tigran V. Shahbazyan}
\affiliation{Department of Physics, Jackson State University, Jackson, MS 39217, USA}
\title{A non-Lorentzian model for strong exciton-plasmon coupling}

\begin{abstract}
We develop a non-Lorentzian analytical model for quantum emitters (QE) resonantly coupled to localized surface plasmons (LSP) in metal-dielectric structures. Using the explicit form of LSP Green function, we derive non-Lorentzian version of semiclassical Maxwell-Bloch equations that describe LSPs directly in terms of metal complex dielectric function rather than via Lorentzian resonances. For a single QE resonantly coupled to an LSP, we obtain an analytical expression for effective optical polarizability of the hybrid system  which, in the Lorentzian approximation, recovers the results of the classical coupled oscillators  model. We demonstrate that non-Lorentzian effects originating from  temporal dispersion of the metal dielectric function affect significantly the optical spectra as the hybrid system transitions to the strong coupling regime. Specifically, in contrast to results of Lorentzian models, the main spectral weight in the system scattering spectra is shifted toward the lower energy polaritonic band, consistent with the experiment.
\end{abstract}
\maketitle


\section{Introduction}
The effects of strong coupling between localized surface plasmons (LSP) in metal-dielectric structures  and quantum emitters (QE) such as excitons in semiconductors or dye molecules have recently attracted considerable interest driven  by numerous potential applications including ultrafast reversible switching \cite{ebbesen-prl11,bachelot-nl13,zheng-nl16}, quantum computing \cite{waks-nnano16,senellart-nnano17} or light harvesting \cite{leggett-nl16}. In the strong coupling regime, coherent energy exchange between QEs and LSP \cite{shahbazyan-nl19,mortensen-rpp20} leads to the emergence of mixed polaritonic states with energy bands separated by the anticrossing gap (Rabi splitting) \cite{novotny-book}. While Rabi splittings in the emission spectra of excitons coupled to  cavity modes in semiconductor microcavities are about several meV  \cite{forchel-nature04,khitrova-nphys06,imamoglu-nature06}, they can reach hundreds  meV in hybrid plasmonic systems involving excitons in \textit{J}-aggregates \cite{bellessa-prl04,sugawara-prl06,wurtz-nl07,fofang-nl08,bellessa-prb09,schlather-nl13,lienau-acsnano14,shegai-prl15}, in various dye molecules \cite{hakala-prl09,berrier-acsnano11,salomon-prl12,luca-apl14,noginov-oe16}  or in semiconductor nanostructures \cite{vasa-prl08,gomez-nl10,gomez-jpcb13,manjavacas-nl11}  resonantly coupled to LSPs. For single QEs, however, reaching   strong coupling regime is a  challenging task as it requires extremely small LSP mode volumes available mainly in nanogaps \cite{hecht-sci-adv19,pelton-sci-adv19,baumberg-natmat2019}.

At the same time, the precise shape of optical spectra in the strong coupling regime  has recently been a subject of active debate \cite{savvidis-aom13,ebbesen-fd15,ebbesen-nc15,shegai-nl17,shegai-acsphot19,zhang-nl17,xu-nl17,garsia-vidal-njp15,ding-prl17,aizpurua-optica18,shahbazyan-nanophot21,xu-acsphot21,shahbazyan-jcp22}. 
In general,  the scattering cross-section for a nanoscale system characterized by localized dipole moment $\mu$ is proportional to $\omega^{4}$,  where  $\omega$ is the incident light frequency \cite{novotny-book}, implying that, in the strong coupling regime, the upper energy polaritonic band should be relatively enhanced. Such a spectral profile originates from a faster dipole radiation rate  at higher frequencies, $\gamma^{\rm rad}=4\mu^{2}\omega^{3}/3\hbar c^{3}$ ($c$ is the speed of light), and it is described, e.g., by the widely-used classical model of two coupled oscillators (CO) \cite{pelton-oe10,pelton-nc18,pelton-ns19}. In the CO model, only one of the oscillators (LSP) interacts with the radiation field while the coupling of the QE to the radiation is neglected due to its much smaller optical dipole moment.  However, recent experiments for excitons resonantly coupled to cavity modes in semiconductor microcavities \cite{savvidis-aom13,ebbesen-fd15,ebbesen-nc15} or to LSPs in metal-dielectric structures \cite{shegai-nl17,zhang-nl17,xu-nl17,shegai-acsphot19,pelton-sci-adv19} reveal the opposite spectral asymmetry pattern with a visible enhancement of the  \textit{lower} energy polaritonic band. For plasmonic systems, a repartitioning of spectral weight between polaritonic bands may arise from the Fano interference between the LSP's dipole moment and the LSP-induced QE's dipole moment \cite{ding-prl17,shahbazyan-nanophot21,xu-acsphot21,shahbazyan-jcp22}. However,  due to a much smaller QE dipole moment, a significant interference effect would require either an extremely strong field confinement \cite{ding-prl17,shahbazyan-nanophot21,xu-acsphot21} or a large number of QEs strongly coupled to the LSP \cite{shahbazyan-jcp22} and, furthermore, would be highly sensitive to the system geometry.

On the other hand, for molecular excitons coupled to a cavity mode,  the accurate spectral weight of polaritonic bands in the emission spectra was obtained within the quantum master equation approach by incorporating the excitation of the  vibronic modes accompanying the emission \cite{garsia-vidal-njp15, aizpurua-optica18}. For plasmonic systems characterized by a frequency-dependent complex dielectric function $\varepsilon (\omega)=\varepsilon'(\omega)+i\varepsilon''(\omega)$ of host metal, the emission spectra with accurate spectral weight distribution have been obtained \cite{shahbazyan-prb22} within the macroscopic quantum electrodynamics approach \cite{welsch-pra98,philbin-njp10}, adopted to metal-dielectric structures supporting LSPs \cite{shahbazyan-prb21}. However, such quantum approaches require extensive numerical efforts that are feasible for specific system geometry and, therefore, are not easily suitable for modeling of experimental optical spectra of hybrid plasmonic systems involving metallic nanostructures of arbitrary and often irregular shape.

In this paper we  present a semiclassical analytical model that fully accounts for temporal dispersion and losses in the metal, encoded in $\varepsilon(\omega)$, and possible interference effects for a single or any number of QEs resonantly coupled to an LSP. Using our recent results for the exact LSP Green function in the quasistatic limit \cite{shahbazyan-pra23}, we develop non-Lorentzian extension of Maxwell-Bloch equations in which the LSP is described directly in terms of metal dielectric function rather than via Lorentzian resonances. In the linear regime, we obtain a formal solution of non-Lorentzian Maxwell-Bloch equations for any number of QEs resonantly coupled to an LSP expressed in terms of bright and dark QE states which is suitable for studying interference effects in such hybrid systems.  For a single QE coupled to a resonant LSP mode, we obtain the system effective optical polarizability, which, in the Lorentzian approximation, recovers the CO model results when only the LSP mode is coupled to the radiation field. By comparing the optical spectra obtained using our non-Lorentzian model and its Lorentzian approximation (CO model), we observe redistribution of the spectral weight toward the lower energy polaritonic band, consistent with the experiment, and trace its origin to temporal dispersion of the metal dielectric function which manifests itself in frequency dependence of the system parameters. Our analytical model can be used for accurate description of experimental spectra of strongly-coupled exciton-plasmon hybrid systems without any significant numerical effort.

%
%
%
%

\section{Non-Lorentzian approach to localized surface plasmons}
\label{sec:lsp}

In this section, we outline our non-Lorentzian approach to localized surface plasmons (LSP) in metal-dielectric structures with characteristic size well below the radiation wavelength \cite{shahbazyan-pra23}. Within quasistatic approach, we define LSP eigenmodes and derive an explicit expression for the LSP Green function which we use to obtain the optical polarizability for the metal nanoparticle (NP) of arbitrary shape in terms of metal dielectric function rather than via Lorentzian LSP resonances. These results will be used in the following sections for setting up non-Lorentzian Maxwell-Bloch equations for QEs resonantly coupled to an LSP and obtaining the effective optical polarizability of hybrid QE-LSP systems.

\subsection{LSP modes and the Green function}


We consider  a metal-dielectric structure  supporting LSP excitations with discrete frequencies $\omega_{n}$ which are localized at a length scale much smaller than the radiation wavelength.  Each region of volume $V_{i}$, metallic or dielectric, is characterized by the dielectric function $\varepsilon_{i}(\omega)$, so that the full dielectric function is $\varepsilon (\omega,\bm{r})=\sum_{i}\theta_{i}(\bm{r})\varepsilon_{i}(\omega)$, where $\theta_{i}(\bm{r})$ is the  unit step function that vanishes outside $V_{i}$. We assume that dielectric regions are characterized by constant permittivities $\varepsilon_{i}$, while for metallic region we adopt complex dielectric function $\varepsilon(\omega)=\varepsilon'(\omega)+i\varepsilon''(\omega)$. In the absence of retardation effects, the LSP modes are defined by the lossless   Gauss equation as \cite{stockman-review}
%
\begin{equation}
\label{gauss-law}
\bm{\nabla}\cdot\left [\varepsilon' (\omega_{n},\bm{r})\bm{\nabla} \Phi_{n}(\bm{r})\right ]=0,
\end{equation}
where $\Phi_{n}(\bm{r})$ and $\bm{E}_{n}(\bm{r})=-\bm{\nabla} \Phi_{n}(\bm{r})$ are, respectively, the potential and electric field of LSP mode $\omega_{n}$ which we chose to be real. Note that LSP eigenmodes are orthogonal in  each  region:
$\int\! dV_{i} \bm{E}_{n}(\bm{r})\cdot\bm{E}_{n'}(\bm{r})=\delta_{nn'}\!\int\! dV_{i} \bm{E}_{n}^{2}(\bm{r})$.

In the presence of metal-dielectric structure, the electromagnetic (EM) dyadic Green function $\bm{D} (\omega;\bm{r},\bm{r}') $ satisfies (in the operator form) 
$\bm{\nabla}\!\times\! \bm{\nabla}\!\times \bm{D}-(\omega^{2}/c^{2})\varepsilon \bm{D} =(4\pi\omega^{2}/c^{2})\bm{I}$,
where $\bm{I}$ is the unit tensor.  The longitudinal part  of $\bm{D}$  is obtained by applying the operator $\bm{\nabla}$ to both sides. In the quasistatic case, the dyadic Green function $\bm{D} (\omega;\bm{r},\bm{r}') $ is related to the scalar Green function for the potentials $D(\omega;\bm{r},\bm{r}')$ as $\bm{D} (\omega;\bm{r},\bm{r}')=\bm{\nabla}\bm{\nabla}'D(\omega;\bm{r},\bm{r}')$. The scalar Green satisfies equation [compare to Eq.~(\ref{gauss-law})]
\begin{equation}
\label{gauss-green-pot}
\bm{\nabla}\cdot\left [\varepsilon (\omega,\bm{r})\bm{\nabla}D(\omega;\bm{r},\bm{r}')\right ]=4\pi \delta(\bm{r}-\bm{r}').
\end{equation}
An explicit expression for the scalar Green function is found by adopting decomposition $D=D_{0}+D_{\rm LSP}$, where $D_{0}(\bm{r}-\bm{r}')=-|\bm{r}-\bm{r}'|^{-1}$ is the free-space Green function and $D_{\rm LSP}(\omega;\bm{r},\bm{r}')$ is the LSP contribution. Expanding that latter over the eigenmodes  of Eq.~(\ref{gauss-law}), we obtain 
\begin{equation}
\label{green-exp}
D_{\rm LSP}(\omega;\bm{r},\bm{r}')=\sum_{n}D_{n}(\omega)\Phi_{n}(\bm{r})\Phi_{n}(\bm{r}'),
\end{equation}
where coefficients $D_{n}(\omega)$ have \textit{non-Lorentzian} form \cite{shahbazyan-pra23}
\begin{equation}
\label{mode-coeff}
D_{n}(\omega)= 
\dfrac{4\pi}{\int\! dV_{\rm m} \bm{E}_{n}^{2}}\frac{1}{\varepsilon' (\omega_{n})-\varepsilon (\omega)}.
\end{equation}
Here, integration takes place over the metal volume $V_{\rm m}$.  Although the expansion in Eq.~(\ref{green-exp}) involves eigenmodes of the lossless Gauss equation (\ref{gauss-law}), the coefficients $D_{n}$ in Eq.~(\ref{mode-coeff}) are defined by the \textit{complex} dielectric function $\varepsilon(\omega)$. 
Accordingly, the non-Lorentzian dyadic Green's function  has the form  $\bm{D}_{\rm LSP} (\omega;\bm{r},\bm{r}')=\sum_{n}D_{n}(\omega)\bm{E}_{n}(\bm{r})\bm{E}_{n}(\bm{r}')$. In the \textit{Lorentzian} approximation, the dielectric function $\varepsilon(\omega)$ in Eq.~(\ref{mode-coeff}) is expanded near the LSP frequencies $\omega_{n}$ as 
\begin{equation}
\label{expand}
\varepsilon(\omega)-\varepsilon' (\omega_{n})=(\omega-\omega_{n})\varepsilon'_{n}+ i\varepsilon''(\omega_{n}),
\end{equation}
where we denoted $\varepsilon'_{n}\equiv \partial\varepsilon' (\omega_{n})/\partial \omega_{n}$. The Lorentzian LSP Green function has the form \cite{shahbazyan-prl16,shahbazyan-prb18}
\begin{equation}
\label{green-lorentzian}
\bm{D}_{\rm LSP}^{L} (\omega;\bm{r},\bm{r}')=\frac{1}{\hbar}\sum_{n}\frac{\tilde{\bm{E}}_{n}(\bm{r})\tilde{\bm{E}}_{n}(\bm{r}')}{\omega_{n}-\omega-i\gamma_{n}/2},
\end{equation}
where 
\begin{equation}
\label{lsp-field-norm}
\tilde{\bm{E}}_{n}(\bm{r})
=\sqrt{\frac{4\pi\hbar}{\varepsilon'_{n}}}
\dfrac{\bm{E}_{n}(\bm{r})}{\left (\int\! dV_{\rm m} \bm{E}_{n}^{2}\right )^{1/2}},
\end{equation}
are normalized LSP mode fields introduced to match the standard Lorentzian expression for the Green function and $\gamma_{n}=2\varepsilon''(\omega_{n})/\varepsilon'_{n}$ is the LSP decay rate. In terms of normalized fields, the LSP optical dipole moment is defined as $\bm{\mu}_{n}=\!\int\! dV\chi' (\omega_{n},\bm{r})\tilde{\bm{E}}_{n}(\bm{r})$, where  $\chi =(\varepsilon-1)/4\pi$ is susceptibility, and the LSP radiative decay rate has the standard form $\gamma_{n}^{\rm rad}=4\mu_{n}^{2}\omega_{n}^{3}/3\hbar c^{3}$.

Finally, the non-Lorentzian LSP Green function represents a sum over the LSP modes $\bm{D}_{\rm LSP} (\omega;\bm{r},\bm{r}')=\sum_{n}\bm{D}_{n} (\omega;\bm{r},\bm{r}')$, where the single-mode Green function
\begin{equation}
\label{lsp-green-norm}
\bm{D}_{n} (\omega;\bm{r},\bm{r}')=
\dfrac{\varepsilon'_{n}}{\hbar}\frac{\tilde{\bm{E}}_{n}(\bm{r})\tilde{\bm{E}}_{n}(\bm{r}')}{\varepsilon' (\omega_{n})-\varepsilon (\omega)}
\end{equation}
has a straightforward Lorentzian limit as $\varepsilon'_{n}$ cancels out the LSP pole residue. We stress that, in contrast to the Lorentzian approximation (\ref{green-lorentzian}), the non-Lorentzian LSP Green function (\ref{lsp-green-norm}) is \textit{exact} for the quasistatic case. Although the above expression for $\bm{D}_{\rm LSP}$ does not explicitly depend on dielectric permittivities  $\varepsilon_{i}$, the latter enter though the LSP frequency $\omega_{n}$ and the LSP mode fields $\tilde{\bm{E}}_{n}(\bm{r})$.

\subsection{Optical polarizability of metal nanoparticles}

With help of the LSP Green function (\ref{lsp-green-norm}), a simple expression for optical polarizability of small NPs of \textit{arbitrary} shape can be obtained \cite{shahbazyan-pra23}. In the following, we consider binary systems, i.e., metal NPs in a dielectric medium with permittivity $\varepsilon_{d}$, which we set $\varepsilon_{d}=1$ for now.  In the presence of incident field $\bm{E}_{0}e^{-i\omega t}$, the field inside the metal is   $\bm{E}(\omega,\bm{r})=\bm{E}_{0}+\bm{E}_{\rm LSP}(\omega,\bm{r})$, where $\bm{E}_{\rm LSP}(\omega,\bm{r})$ is the LSP-induced field given by
\begin{align}
\bm{E}_{\rm LSP}(\omega,\bm{r})=\int dV'\bm{D}_{\rm LSP} (\omega;\bm{r},\bm{r}')\chi(\omega,\bm{r}')\bm{E}_{0}
\nonumber\\
=\sum_{n}c_{n}\tilde{\bm{E}}_{n}(\bm{r})\frac{\varepsilon (\omega)-1}{\varepsilon' (\omega_{n})-\varepsilon (\omega)},
\end{align}
where, using Eq.~(\ref{lsp-green-norm}), the coefficients $c_{n}$ are found as $c_{n}=(\varepsilon'_{n}/4\pi\hbar)\int dV_{\rm m}\tilde{\bm{E}}_{n}(\bm{r})\cdot \bm{E}_{0}$ and we used that $\chi(\omega,\bm{r}')=[\varepsilon(\omega)-1]/4\pi$ inside the metal and vanishes outside of it. Note now that the LSP mode fields $\tilde{\bm{E}}_{n}(\bm{r})$ are regular inside the metal and therefore, using Eq.~(\ref{lsp-field-norm}), the external field can be expanded as $\bm{E}_{0}=\sum_{n}c_{n}\tilde{\bm{E}}_{n}(\bm{r})$ with the same coefficients $c_{n}$. Then the field inside the metal takes the form
\begin{equation}
\bm{E}(\omega,\bm{r})=\sum_{n}c_{n}\tilde{\bm{E}}_{n}(\bm{r})\frac{\varepsilon' (\omega_{n})-1}{\varepsilon' (\omega_{n})-\varepsilon (\omega)}.
\end{equation}
Using the above expression, the \textit{induced} LSP dipole moment  $\bm{p}=\int dV_{\rm m}\chi(\omega)\bm{E}(\omega,\bm{r})$ can be presented as $\bm{p}(\omega)=\sum_{n}\bm{p}_{n}(\omega)$, where 
\begin{equation}
\label{lsp-response}
\bm{p}_{n}(\omega)=\dfrac{\varepsilon'_{n}}{\hbar}\frac{\bm{\mu}_{n}(\omega)\,\bm{\mu}_{n}\!\cdot\! \bm{E}_{0}}{\varepsilon' (\omega_{n})-\varepsilon (\omega)},
\end{equation}
is the induced dipole moment of LSP mode. Here, $\bm{\mu}_{n}(\omega)=\chi (\omega)\int dV_{\rm m}\tilde{\bm{E}}_{n}(\bm{r})$ is LSP mode's \textit{frequency-dependent} optical dipole and $\bm{\mu}_{n}\equiv\bm{\mu}_{n}(\omega_{n})$ is its value at the LSP frequency. The induced LSP dipole moment (\ref{lsp-response}) defines the  LSP optical polarizability tensor $\bm{\alpha}_{n}(\omega)$ via the standard relation $\bm{p}_{n}(\omega)=\bm{\alpha}_{n}(\omega)\bm{E}_{0}$ with 
\begin{equation}
\label{pol-lsp}
\bm{\alpha}_{n}(\omega)=\dfrac{\varepsilon'_{n}}{\hbar}\frac{\bm{\mu}_{n}(\omega)\,\bm{\mu}_{n}}{\varepsilon' (\omega_{n})-\varepsilon (\omega)}.
\end{equation}
To present the LSP polarizability  in a more symmetric form, we use Eq.~(\ref{lsp-field-norm}) to express it via original LSP mode fields. Then Eq.~(\ref{pol-lsp}) can be recast  as  $\bm{\alpha}_{n}(\omega)=\alpha_{n}(\omega)\bm{e}_{n}\bm{e}_{n}$, where 
 \begin{equation}
 \label{pol-small}
 \alpha_{n}(\omega)=V_{n}\, \dfrac{\varepsilon(\omega)-1}{\varepsilon(\omega)-\varepsilon'(\omega_{n})},
 \end{equation}
is scalar polarizability, $\bm{e}_{n}=\int\! dV_{\rm m}\bm{E}_{n}/|\int\! dV_{\rm m}\bm{E}_{n}|$ is unit vector for  LSP mode polarization, $V_{n}$ is \textit{effective} system volume  defined as
\begin{equation}
\label{Vn}
V_{n}=V_{\rm m}|\chi'(\omega_{n})|s_{n},
~~~
s_{n}=\dfrac{\left (\int\! dV_{\rm m}\bm{E}_{n}\right )^{2}}{V_{\rm m}\int\! dV_{\rm m} \bm{E}_{n}^{2}},
\end{equation}
and  $s_{n}\leq 1$ is parameter depending on system geometry ($s_{n}=1$ for spherical and spheroidal NPs) \cite{shahbazyan-pra23}. Note that for a spherical NP of radius $a$, we have $\varepsilon'(\omega_{n})=-2$ and hence $V_{n}=a^{3}$, recovering the standard expression for its polarizability.

The non-Lorentzian LSP polarizability (\ref{pol-small}) is \textit{exact} in the quasistatic limit for NPs of any shape. For larger NPs beyond the quasistatic limit, the LSP radiation damping is included in the standard way via replacement \cite{novotny-book} $ \alpha_{n} \rightarrow \alpha_{n}\left [ 1- (2i/3)k^{3}\alpha_{n}\right]^{-1}$. Restoring the medium dielectric constant $\varepsilon_{d}$, we finally obtain
 \begin{equation}
 \label{pol-small2}
\alpha_{n}(\omega)=\dfrac{V_{n}[\varepsilon(\omega)-\varepsilon_{d}]}{\varepsilon(\omega)\!-\!\varepsilon'(\omega_{n})\!-\!\frac{2i}{3}k^{3}V_{n}[\varepsilon(\omega)\!-\!\varepsilon_{d}]},
 \end{equation}
where $V_{n}=V_{\rm m}|\varepsilon'(\omega_{n})/\varepsilon_{d}-1|s_{n}/4\pi$. In the Lorentzian approximation, expanding $\varepsilon(\omega)$ near $\omega_{n}$ according Eq.~(\ref{expand}) and using again Eq.~(\ref{lsp-field-norm}), we recover the standard expression for polarizability tensor of LSP treated as a localized dipole
\begin{equation}
 \label{pol-L-small2}
\bm{\alpha}_{n}^{L}(\omega)=\frac{1}{\hbar}\frac{\bm{\mu}_{n}\bm{\mu}_{n}}{\omega_{n}-\omega-i\gamma_{n}/2},
\end{equation}
where the LSP decay rate now includes both non-radiative and radiative processes: $\gamma_{n}=2\varepsilon'' (\omega_{n})/\varepsilon'_{n}+4\mu_{n}^{2}\omega_{n}^{3}/3\hbar c^{3}$. In terms of LSP polarizability, the extinction and scattering cross-sections are given by
\begin{equation}
\label{cross-sections}
\sigma_{\rm ext}(\omega)\!=\frac{4\pi \omega}{c}\,{\rm Im}\, \alpha_{n}(\omega),
~
\sigma_{\rm scatt}(\omega)\!=\frac{8\pi \omega^{4}}{3c^{4}}\,|\alpha_{n}(\omega)|^{2}.
\end{equation}
%

%
\begin{figure}[tb]
\vspace{2mm}
\centering
\includegraphics[width=0.99\columnwidth]{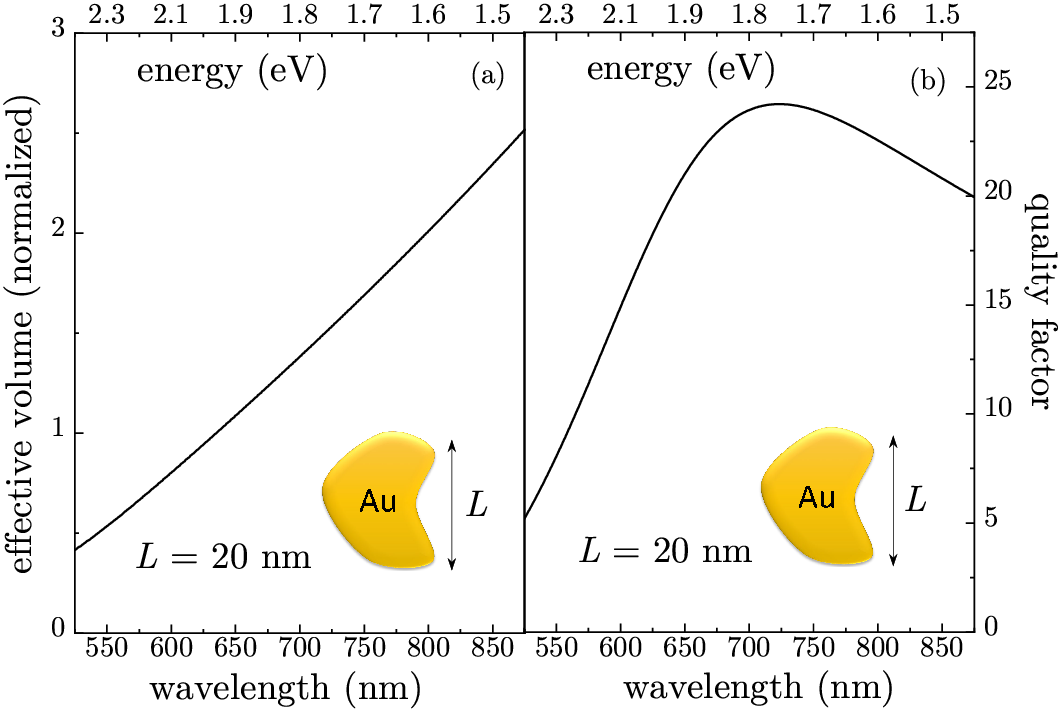}
\caption{\label{fig1} Normalized LSP effective volume $V_{n}/V_{m}$ (a) and quality factor $Q_{n}=\omega_{n}/\gamma_{n}$ (b) for an Au NP in water are plotted against the LSP wavelength. Inset: Schematics of Au NP of irregular shape with characteristic size $L=20$ nm.
 }
\vspace{-4mm}
\end{figure}
%

\begin{figure}[tb]
\vspace{2mm}
\centering
\includegraphics[width=0.99\columnwidth]{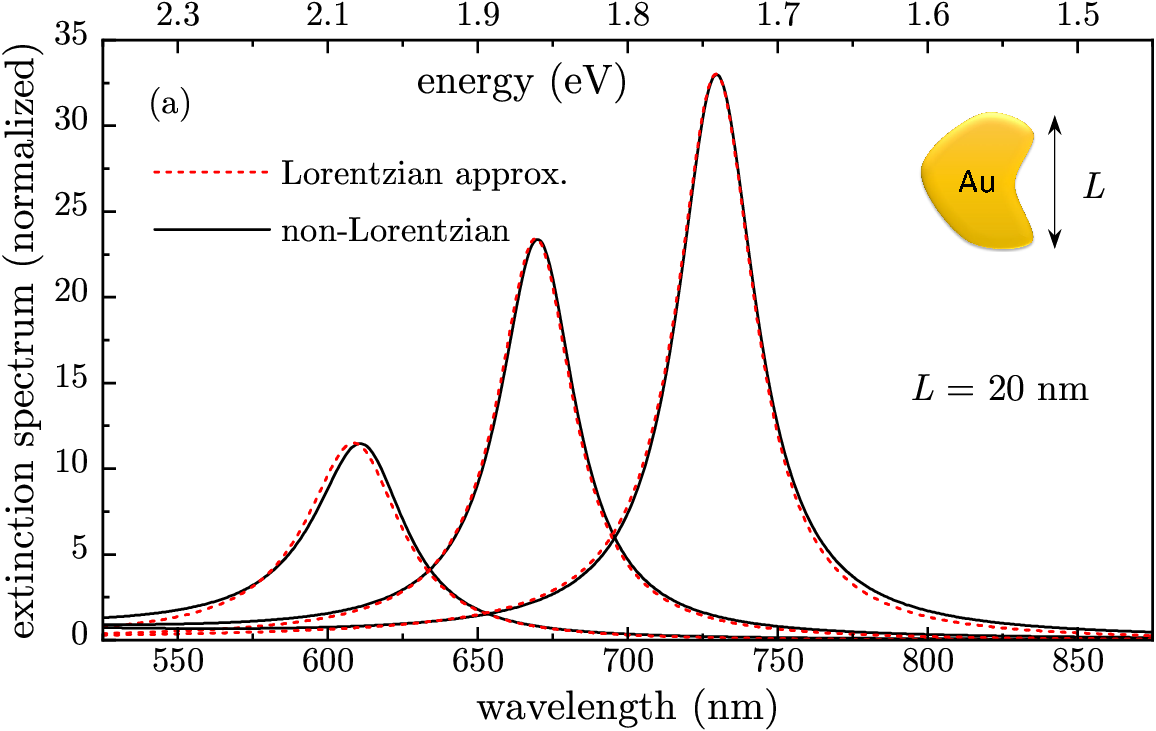}

\vspace{3mm}

\includegraphics[width=0.99\columnwidth]{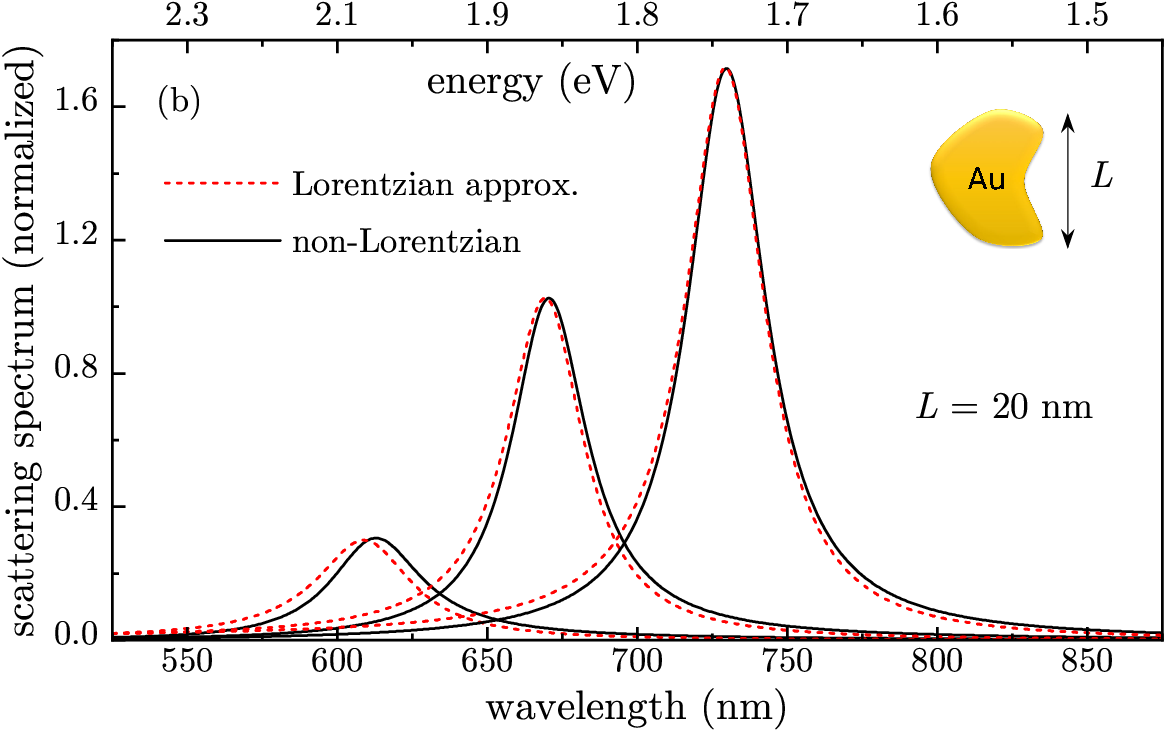}
\caption{\label{fig2} Normalized extinction cross-section $\sigma_{\rm ext}/L^{2}$ (a) and scattering cross-section $\sigma_{\rm scatt}/L^{2}$ (b) for an Au NP in water at LSP wavelengths 610 nm, 670 nm, and 730 nm calculated using non-Lorentzian and Lorentzian models. 
 }
\vspace{-4mm}
\end{figure}

The effect of temporal dispersion of metal dielectric function on the optical polarizability is illustrated in Figs.~\ref{fig1} and \ref{fig2} for an Au NP placed in water ($\varepsilon_{d}=1.77$). We consider Au NP \textit{without} specific shape but with characteristic size $L=20$ nm and the metal volume $V_{\rm m}=L^{3}$.  We use the \textit{experimental} Au dielectric function in all calculations. For simplicity, we  assume that incident light polarization is aligned with that of LSP and set $s_{n}=1$ hereafter. In Fig.~(\ref{fig1}), we plot the normalized effective volume $V_{n}/V_{\rm m}$ and the LSP quality factor $Q_{n}=\omega_{n}/\gamma_{n}$ versus the LSP wavelength.  The effective volume, which is determined by $\varepsilon'(\omega_{n})$, increases monotonically with the LSP wavelength, while the quality factor first increases  but exhibits a maximum at about 700 nm due to non-monotonic behavior of $\varepsilon'' (\omega_{n})$. 

The frequency dependence of system parameters determines the amplitude and width of LSP-dominated extinction and scattering spectra shown in Fig.~(\ref{fig2}). The spectra were calculated for several typical LSP wavelengths $\lambda_{n}= 610$ nm, 670 nm, and 730 nm using non-Lorentzian polarizability (\ref{pol-small2}) and its Lorentzian approximation (\ref{pol-L-small2}). In fact, the Lorentzian approximation is largely accurate except  a noticeable blueshift in the shorter wavelength region. Later in this paper we demonstrate that, in the presence of QEs resonantly coupled to the LSP, non-Lorentzian effects lead to significant changes in the optical spectra as the system transitions to strong coupling regime.

%
%

\section{Non-Lorentzian Maxwell-Bloch equations}
\label{sec:mb}


In this section, we use the approach developed in the previous section to set up non-Lorentzian Maxwell-Bloch (MB) equations describing optical interactions of $N$ QEs with a single resonant LSP mode. Each QE is characterized by optical polarizability
\begin{equation}
\label{polar-qe}
\bm{\alpha}_{i}(\omega)=\frac{1}{\hbar}\frac{\bm{\mu}_{i}\bm{\mu}_{i}}{\omega_{i}-\omega-\frac{i}{2}\gamma_{i}},
\end{equation}
where $\omega_{i}$ is QE's excitation frequency that is close to resonance with the LSP frequency $\omega_{n}$, $\gamma_{i}$ is the linewidth, and $\bm{\mu}_{i}=\mu_{i}\bm{n}_{i}$ is the dipole moment ($\bm{n}_{i}$ is its orientation). In the presence of incident light $\bm{E}_{0}e^{-i\omega t}$, the induced dipole moment of LSP mode has the form 
\begin{equation}
\label{lsp-dipole-ind}
\bm{p}_{n}=\dfrac{\varepsilon'_{n}}{\hbar}\frac{\bm{\mu}_{n}(\omega)\,\bm{\mu}_{n}\!\cdot\! \bm{E}_{0}}{\varepsilon' (\omega_{n})-\varepsilon (\omega)}+\bm{p}_{n}^{\rm qe},
\end{equation}
where the first term is induced by the incident field [compare to Eq.~(\ref{lsp-response})] and the second term $\bm{p}_{n}^{\rm qe}=\int dV_{\rm m}\chi(\omega)\bm{E}_{\rm qe}$ is  induced by the QEs' electric field
\begin{equation}
\label{field-emitters}
\bm{E}_{\rm qe}(\bm{r})=\sum_{i}\bm{D}_{n} (\omega;\bm{r},\bm{r}_{i})\bm{p}_{i}
=\sum_{i}\dfrac{\varepsilon'_{n}}{\hbar}\frac{\tilde{\bm{E}}_{n}(\bm{r})\tilde{\bm{E}}_{n}(\bm{r}_{i})\!\cdot\! \bm{p}_{i}}{\varepsilon' (\omega_{n})-\varepsilon (\omega)}.
\end{equation}
Here, $\bm{p}_{i}$ are QEs' induced dipole moments and $\bm{D}_{n} (\omega;\bm{r},\bm{r}_{i})$ is single-mode LSP Green function (\ref{lsp-green-norm}). Introducing QE's polarizations $\rho_{i}$ defined as $\bm{p}_{i}=\bm{\mu}_{i}\rho_{i}$, the QE-induced LSP dipole moment takes the form
\begin{equation}
\bm{p}_{n}^{\rm qe}=\frac{\varepsilon'_{n}\bm{\mu}_{n}(\omega)}{\varepsilon' (\omega_{n})-\varepsilon (\omega)}\sum_{i}g_{in}\rho_{i}
\end{equation}
where $g_{in}=\bm{\mu}_{i}\cdot\tilde{\bm{E}}_{n}(\bm{r}_{i})/\hbar$ is  QE-LSP \textit{coupling} parameter. Then the induced LSP dipole moment (\ref{lsp-dipole-ind}) takes the form
\begin{equation}
\bm{p}_{n}=\frac{\varepsilon'_{n}\bm{\mu}_{n}(\omega)}{\varepsilon' (\omega_{n})-\varepsilon (\omega)}\left (\sum_{i}\rho_{i}g_{in}+\bm{\mu}_{n}\!\cdot\! \bm{E}_{0}/\hbar\right ),
\end{equation}
Finally, defining LSP polarization $\rho_{n}$ through the relation $\bm{p}_{n}=\bm{\mu}_{n}(\omega)\rho_{n}$, we obtain the first MB equation as
\begin{equation}
\label{mb-lsp}
\Omega_{n}\rho_{n}=\sum_{i}\rho_{i}g_{in}+\bm{\mu}_{n}\!\cdot\! \bm{E}_{0}/\hbar,
\end{equation}
where we introduced non-Lorentzian detuning
%
\begin{equation}
\label{detuning}
\Omega_{n}(\omega)=[\varepsilon' (\omega_{n})-\varepsilon (\omega)]/\varepsilon'_{n}.
\end{equation}
In the Lorentzian limit, by expanding $\varepsilon (\omega)$ near $\omega_{n}$ with help of Eq.~(\ref{expand}), we obtain $\Omega_{n}^{L}=\omega_{n}-\omega-i\gamma_{n}/2$, recovering standard MB equation \cite{shahbazyan-jcp22} for the LSP polarization $\rho_{n}$. The second MB equation for QEs' polarization has the standard form
\begin{equation}
\label{mb-emitter}
\Omega_{i}\rho_{i}=g_{in}\rho_{n}+\bm{\mu}_{i}\!\cdot\! \bm{E}_{0}/\hbar,
\end{equation}
where $\Omega_{i}(\omega)=\omega_{i}-\omega-\frac{i}{2}\gamma_{i}$. 

The coupled system of Eqs.~(\ref{mb-lsp}) and (\ref{mb-emitter}) represents non-Lorentzian extension of coupled MB equations for polarizations $\rho_{n}$ and $\rho_{i}$ as $\Omega_{n}(\omega)$ now incorporates full complex metal dielectric function $\varepsilon (\omega)$. Importantly, the LSP dipole moment  $\bm{p}_{n}(\omega)=\bm{\mu}_{n}(\omega)\rho_{n}$ has additional dependence on $\varepsilon (\omega)$ via $\bm{\mu}_{n}(\omega)=[\varepsilon (\omega)-1]\int dV_{\rm m}\tilde{\bm{E}}_{n}(\bm{r})/4\pi$. Note that the QE-LSP coupling parameter $g_{in}=\bm{\mu}_{i}\cdot\tilde{\bm{E}}_{n}(\bm{r}_{i})/\hbar$ is \textit{independent} of frequency and, using Eq.~(\ref{lsp-field-norm}), can be recast in a cavity-like form as \cite{shahbazyan-nl19}
\begin{equation}
\label{coupling-mode-volume}
g^{2}_{in}
=\frac{2\pi \mu_{i}^{2}\omega_{n}}{\hbar{\cal V}^{(i)}_{n}},
~~
\frac{1}{{\cal V}^{(i)}_{n}}
= \frac{2[\bm{n}_{i}\!\cdot\!\bm{E}_{n}(\bm{r}_{i})]^{2}}{\omega_{n}\varepsilon'_{n}\int \! dV_{\rm m} \bm{E}_{n}^{2}},
\end{equation}
where ${\cal V}^{(i)}_{n}$ is  projected  LSP mode volume that characterizes  the LSP field confinement at a point $\bm{r}_{i}$ in the direction $\bm{n}_{i}$ \cite{shahbazyan-prl16,shahbazyan-acsphot17,shahbazyan-prb18}. 


In the linear regime, the non-Lorentzian MB equations can be straightforwardly solved. First, it follows from Eq.~(\ref{mb-emitter}) that
\begin{equation}
\label{sum}
\sum_{i} \rho_{i}g_{in}=\Sigma_{n}(\omega)\rho_{n}+\bm{q}_{n}(\omega)\!\cdot\! \bm{E}_{0}/\hbar.
\end{equation}
Here, $\bm{q}_{n}(\omega)=\sum_{i}\bm{q}_{in}(\omega)$, where $\bm{q}_{in}(\omega)=g_{in}\bm{\mu}_{i}/\Omega_{i}=
\bm{\alpha}_{i}(\omega)\tilde{\bm{E}}_{n}(\bm{r}_{i})$ is QE's dipole moment \textit{induced} by the LSP mode field, while 
\begin{equation}
\label{self}
\Sigma_{n}(\omega)=\frac{1}{\hbar}\sum_{i}\bm{q}_{in}(\omega)\!\cdot\! \tilde{\bm{E}}_{n}(\bm{r}_{i})
=\sum_{i}\frac{g_{in}^{2}}{\Omega_{i}(\omega)},
\end{equation}
is LSP's self-energy due to its coupling to QEs. After substituting Eq.~(\ref{sum}) into Eq.~(\ref{mb-lsp}), the LSP MB equation takes the form $[\Omega_{n}(\omega)-\Sigma_{n}(\omega)]\rho_{n}=[\bm{\mu}_{n}+\bm{q}_{n}(\omega)]\!\cdot\! \bm{E}_{0}/\hbar$, and we obtain the LSP induced dipole moment $\bm{p}_{n}(\omega)=\bm{\mu}_{n}(\omega)\rho_{n}$ as 
\begin{equation}
\label{dipole-induced-lsp}
\bm{p}_{n}(\omega)=\frac{\bm{\mu}_{n}(\omega)}{\hbar}\frac{[\bm{\mu}_{n}+\bm{q}_{n}(\omega)]\!\cdot\! \bm{E}_{0}}{\Omega_{n}(\omega)-\Sigma_{n}(\omega)}.
\end{equation}
From the QE MB equation (\ref{mb-emitter}), the induced QE  dipole moment has the form $\bm{p}_{i}=\bm{\mu}_{i}\rho_{i}=\bm{q}_{n}(\omega)\rho_{n}+\bm{\alpha}_{i}(\omega)\bm{E}_{0}$, and so the QEs' combined dipole moment $\bm{p}_{\rm qe}=\sum_{i}\bm{p}_{i}$ is obtained as
\begin{equation}
\label{dipole-induced-em}
\bm{p}_{\rm qe}(\omega)=\frac{\bm{q}_{n}(\omega)}{\hbar}\frac{[\bm{\mu}_{n}+\bm{q}_{n}(\omega)]\!\cdot\! \bm{E}_{0}}{\Omega_{n}(\omega)-\Sigma_{n}(\omega)} +\bm{\alpha}_{N}(\omega)\bm{E}_{0}.
\end{equation}
where $\bm{\alpha}_{N}(\omega)=\sum_{i}\bm{\alpha}_{i}(\omega)$. The combination $\bm{\mu}_{n}+\bm{q}_{n}$ in the numerator of Eqs.~(\ref{dipole-induced-lsp}) and (\ref{dipole-induced-em}) indicates that the LSP optical dipole is now enhanced by LSP-mode-induced dipole of QEs excited \textit{in-phase} by the LSP near field. 

The above expressions for dipole moments greatly simplify if QEs are characterized by a single excitation frequency $\omega_{i}=\omega_{0}$ and decay rate $\gamma_{i}=\gamma_{0}$. In this case, the LSP self-energy takes the form $\Sigma_{n}(\omega)=g^{2}/\Omega_{0}(\omega)$, where
\begin{equation}
\label{coupling-N}
g^{2}=\sum_{i}g_{in}^{2},
~~~
\Omega_{0}(\omega)=\omega_{0}-\omega-i\frac{\gamma_{0}}{2},
\end{equation}
while QEs form bright and dark collective states, the former strongly coupled to LSP and the latter not coupled at all. Introducing the dipole moment of bright QE state as \cite{shahbazyan-jcp22} $\bm{\mu}_{\rm b}=g^{-1}\sum_{i}g_{in}\bm{\mu}_{i}$, the LSP-mode-induced dipole moment of QE ensemble can be presented as $\bm{q}_{n}(\omega)=g\bm{\mu}_{\rm b}/\Omega_{0}$, and we obtain after some algebra
\begin{equation}
\label{qe-dipole}
\bm{p}_{\rm qe}(\omega)=\bm{\alpha}_{\rm d}(\omega)\bm{E}_{0}
+\frac{1}{\hbar}\frac{\bm{\mu}_{\rm b}(\bm{\mu}_{\rm b}\!\cdot\! \bm{E}_{0})}{\Omega_{0}-g^{2}/\Omega_{n}}
+\frac{1}{\hbar}\frac{\bm{q}_{n}(\omega)(\bm{\mu}_{n}\!\cdot\! \bm{E}_{0})}{\Omega_{n}-g^{2}/\Omega_{0}}.
\end{equation}
Here, the first term represents contribution of dark states characterized by polarizability $\bm{\alpha}_{\rm d}=\bm{\alpha}_{N}-\bm{\mu}_{\rm b}\bm{\mu}_{\rm b}/\hbar\Omega_{0}$, second term is bright state contribution and last cross term describes QE-LSP interference. A similar interference term also  appears in the LSP innduced dipole moment, which now has the form
\begin{equation}
\label{lsp-dipole}
\bm{p}_{n}(\omega)=\frac{\bm{\mu}_{n}(\omega)}{\hbar}\frac{[\bm{\mu}_{n}+\bm{q}_{n}(\omega)]\!\cdot\! \bm{E}_{0}}{\Omega_{n}(\omega)-g^{2}/\Omega_{0}}.
\end{equation}
The system full dipole moment is $\bm{p}(\omega)=\bm{p}_{n}(\omega)+\bm{p}_{e}(\omega)$ that defines
the  system's effective optical polarizability. The cross terms in Eqs.~(\ref{qe-dipole}) and (\ref{lsp-dipole}) give rise to Fano interference between the bright collective QE state and plasmonic antenna, which is described by the first term in Eq.~(\ref{lsp-dipole}) \cite{shahbazyan-nanophot21,shahbazyan-jcp22}. Note, however, that due to a much larger LSP optical dipole moment, the Fano interference effects are important only for large QE numbers.



\section{Effective polarizability and optical spectra}
\label{sec:single}

In this section, we obtain explicit expression for effective optical polarizability of a \textit{single} QE with dipole moment $\bm{\mu}_{0}$ which is  strongly coupled to a resonant LSP mode. For a single QE, we have $\bm{q}_{n}=g\bm{\mu}_{0}/\Omega_{0}(\omega)$ and so the induced dipole moments (\ref{qe-dipole}) and (\ref{lsp-dipole}) take, respectively, the form
\begin{equation}
\label{dipole-induced-single-em}
\bm{p}_{\rm qe}(\omega)=\frac{\bm{\mu}_{0}}{\hbar}\frac{[\bm{\mu}_{0}+g\bm{\mu}_{n}/\Omega_{n}(\omega)]\!\cdot\! \bm{E}_{0}}{\Omega_{0}(\omega)-g^{2}/\Omega_{n}(\omega)}
\end{equation}
and
\begin{equation}
\label{dipole-induced-single-lsp}
\bm{p}_{n}(\omega)=\frac{\bm{\mu}_{n}(\omega)}{\hbar}\frac{[\bm{\mu}_{n}+g\bm{\mu}_{0}/\Omega_{0}(\omega)]\!\cdot\! \bm{E}_{0}}{\Omega_{n}(\omega)-g^{2}/\Omega_{0}(\omega)}.
\end{equation}
In Eq.~(\ref{dipole-induced-single-em}), the first term represents contribution of the QE coupled to the plasmonic antenna, while the second term describes interference effect when the light is first absorbed by the antenna  and then re-emitted by the LSP-mode-induced QE dipole moment  $\bm{q}_{n}(\omega)$. In Eq.~(\ref{dipole-induced-single-lsp}), the first term represents contribution of the plasmonic antenna coupled to the emitter, while the second term describes interference effect when the light is first absorbed by the LSP-mode-induced QE dipole moment $\bm{q}_{n}(\omega)$ and then re-emitted by the antenna.

In the following, we assume, for simplicity, that LSP's and QE's dipoles are aligned with the incident field polarization, i.e., $\bm{E}_{0}\parallel \bm{\mu}_{n}\parallel \bm{\mu}_{0}$. In this case, we have $\bm{p}_{n}(\omega)=\tilde{\alpha}_{n}(\omega)\bm{E}_{0}$ and $\bm{p}_{\rm qe}(\omega)=\tilde{\alpha}_{\rm qe}(\omega)\bm{E}_{0}$ for induced moments (\ref{dipole-induced-single-lsp}) and (\ref{dipole-induced-single-em}), respectively, where $\tilde{\alpha}_{n}(\omega)$ and $\tilde{\alpha}_{\rm qe}(\omega)$ are the corresponding scalar effective polarizabilities.  The effective LSP polarizability $\tilde{\alpha}_{n}(\omega)$ has the form
\begin{equation}
\label{pol-antenna}
\tilde{\alpha}_{n}(\omega)=\frac{1}{\hbar}\frac{\mu_{n}(\omega)\mu_{n}f_{n}(\omega)}{\Omega_{n}(\omega)-g^{2}/\Omega_{0}(\omega)}
\end{equation}
where the function $f_{n}(\omega)=1+g\mu_{0}/\mu_{n}\Omega_{0}(\omega)$ describes Fano interference in the scattering spectra between the plasmonic antenna and LSP-mode-induced QE dipole. Since the LSP dipole moment is much larger than that of QE, $\mu_{n}/\mu_{i}\gg 1$, for a single QE the Fano interference effects are relatively weak ($f_{n}\approx 1$) although they can be significant for many QEs \cite{shahbazyan-nanophot21,shahbazyan-jcp22}. At the same time, the effective QE polarizability has the form
\begin{equation}
\label{pol-emitter}
\tilde{\alpha}_{\rm qe}(\omega)=\frac{1}{\hbar}\frac{\mu_{0}^{2}f_{0}(\omega)}{\Omega_{0}(\omega)-g^{2}/\Omega_{n}(\omega)},
\end{equation}
where the function $f_{0}(\omega)=1+g\mu_{n}/\mu_{0}\Omega_{n}(\omega)$ describes Fano interference between the LSP-mode-induced QE dipole and plasmonic antenna. Since $\mu_{n}/\mu_{0}\gg 1$, the interference contribution to $\tilde{\alpha}_{\rm qe}$ is dominant ($f_{\rm qe}\gg 1$), in contrast to  $\tilde{\alpha}_{n}$. Note, however, that both interference contributions  are of the same order of magnitude while being much smaller than that of plasmonic antenna. 

The effective polarizability of the QE-LSP hybrid system  is $\tilde{\alpha}(\omega)=\tilde{\alpha}_{n}(\omega)+\tilde{\alpha}_{\rm qe}(\omega)$, which defines the extinction and scattering cross-sections according to Eq.~(\ref{cross-sections}).  The system radiative damping can be included by the  replacement $\tilde{\alpha} \rightarrow \tilde{\alpha}\left [ 1- (2i/3)k^{3}\tilde{\alpha}\right]^{-1}$, similar to NP polarizability (\ref{pol-small2}). The Lorenntzian approximation is obtained by replacing $\Omega_{n}(\omega)=[\varepsilon' (\omega_{n})-\varepsilon (\omega)]/\varepsilon'_{n}$ with $\Omega_{n}^{L}(\omega)=\omega_{n}-\omega-i\gamma_{n}/2$ and $\mu_{n}(\omega)$ with $\mu_{n}(\omega_{n})\equiv \mu_{n}$ in Eqs.~(\ref{pol-antenna}) and (\ref{pol-emitter}).

For a single QE, the QE's coupling to radiation is negligibly small as compared to that of LSP, so the main contribution to  $\tilde{\alpha}(\omega)$  comes from the plasmonic antenna contribution Eq.~(\ref{pol-antenna}) (with $f_{n}=1$), which, using Eq.~(\ref{lsp-field-norm}), can be recast as [compare to Eq.~(\ref{pol-small2})]
 \begin{equation}
 \label{pol-antenna2}
 \tilde{\alpha}_{n}(\omega)=\dfrac{V_{n}[\varepsilon(\omega)-\varepsilon_{d}]}{\varepsilon(\omega)-\varepsilon'(\omega_{n})\!-\!\frac{2i}{3}k^{3}V_{n}[\varepsilon(\omega)\!-\!\varepsilon_{d}]-\frac{\varepsilon'_{n}g^{2}}{\omega-\omega_{0}+i\gamma_{0}/2}}.
 \end{equation}
In the Lorentzian approximation, the above expression reduces to [compare to Eq.~(\ref{pol-L-small2})]
\begin{equation}
 \label{pol-L-antenna2}
\tilde{\alpha}_{n}^{L}(\omega)=\frac{\mu_{n}^{2}}{\hbar}\frac{\omega_{0}-\omega-i\gamma_{0}/2}{(\omega_{n}-\omega-i\gamma_{n}/2)(\omega_{0}-\omega-i\gamma_{0}/2)-g^{2}},
\end{equation}
which coincides with the effective polarizability obtained within CO model \cite{pelton-oe10}. In the following section,  show that non-Lorentzian effects strongly affect the  optical spectra as the system transitions to strong coupling regime.

\section{Discussion and numerical results}
\label{sec:num}

Here we illustrate non-Lorentzian effects in the transition to strong coupling regime for a QE situated near an Au NP in water ($\varepsilon_{d}=1.77$) with excitation frequency in resonance with LSP frequencies, i.e., $\omega_{0}=\omega_{n}$, considered here as \textit{input parameters}.  The NP characteristic size $L$, which defines its volume as $V_{\rm m}=L^{3}$,  is chosen $L=20$ nm, and the experimental Au dielectric function is used in all calculations. The QE dipole moment, LSP polarization and incident light polarization are all aligned. The ratio of QE and LSP dipole moments is taken $\mu_{0}/\mu_{n}=10^{-4}$ while the ratio of QE and LSP spectral widths is  $\gamma_{0}/\gamma_{n}=0.2$. For a single QE, the largest by far contribution come from the antenna's polarizability (\ref{pol-antenna2}) and its Lorentzian counterpart (\ref{pol-L-antenna2}) although all contributions are included in the numerical calculations.

%
\begin{figure}[tb]
\begin{center}
\includegraphics[width=0.99\columnwidth]{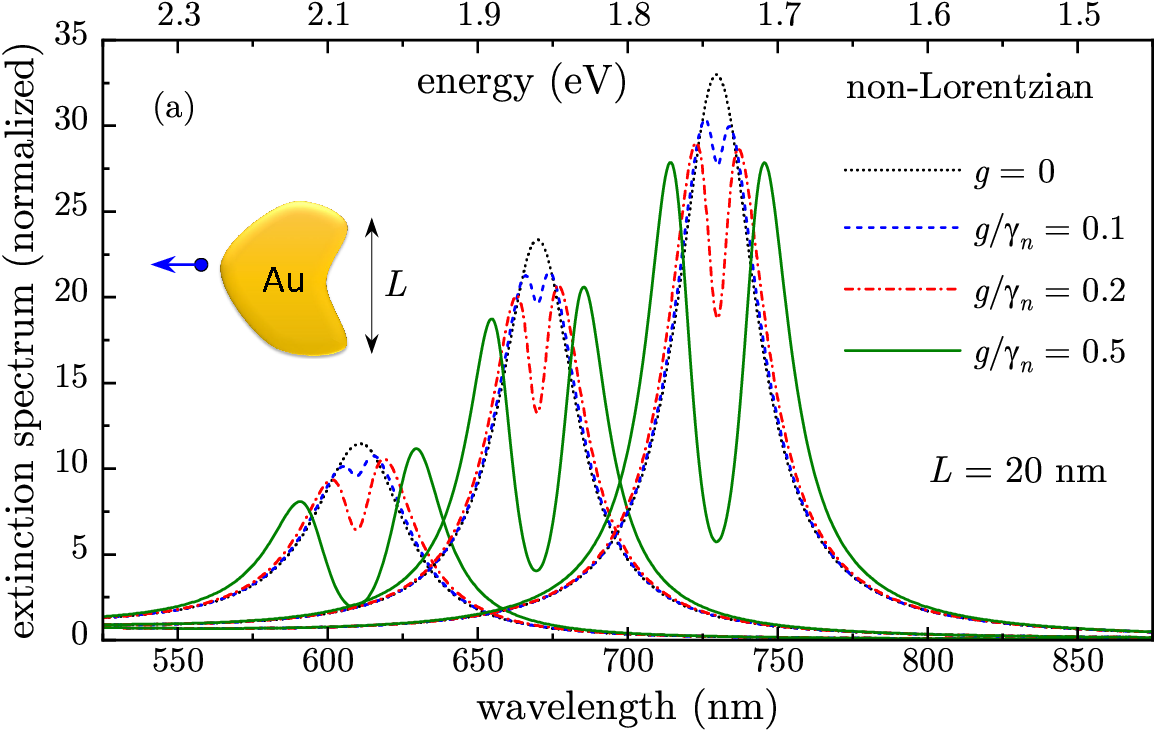}

\vspace{3mm}

\includegraphics[width=0.99\columnwidth]{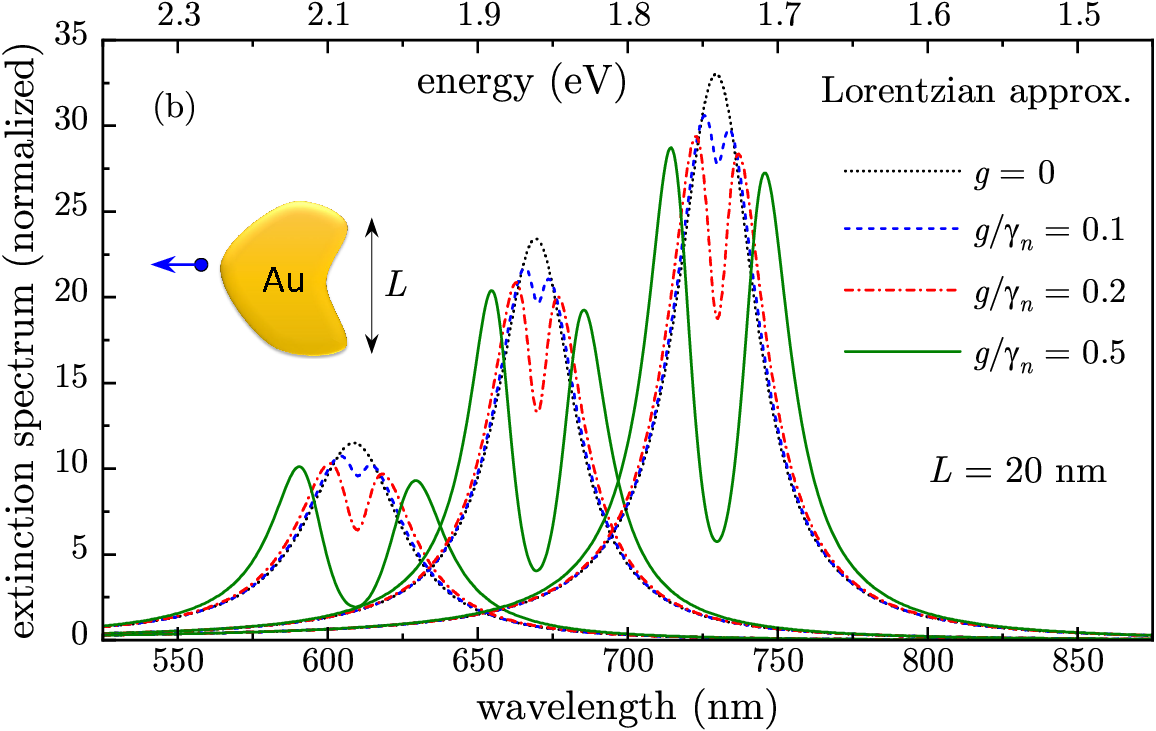}
\caption{\label{fig3} Normalized extinction cross-section $\sigma_{\rm ext}/L^{2}$ calculated using non-Lorentzian model (a) and its Lorentzian approximation (b) for a QE near Au NP in water at LSP wavelengths 610 nm, 670 nm, and 730 nm. Inset: Schematics of a QE near Au NP of irregular shape.
 }
\end{center}
\vspace{-4mm}
\end{figure}
%

In Fig.~\ref{fig3} we show the normalized extinction spectra $\sigma_{\rm ext}/L^{2}$ calculated using our non-Lorentzian model [see Fig.~\ref{fig3}(a)] and its Lorentzian approximation [see Fig.~(\ref{fig3}(b)] for typical LSP wavelengths 610 nm, 670 nm and 730 nm. Although numerical calculations were performed using full effective polarizability, the dominant contribution comes from plasmonic antenna given by Eq.~(\ref{pol-antenna2}) (for non-Lorenzian) and Eq.~(\ref{pol-L-antenna2}) (for Lorentzian), so that Fig.~\ref{fig3}(a) and Fig.~\ref{fig3}(b) provide, in fact, direct comparison between our non-Lorenzian model and classical CO model. With increasing ratio $g/\gamma_{n}$, the spectra first develop a narrow minimum corresponding to exciton-induced transparency (ExIT) \cite{pelton-oe10,pelton-nc18,pelton-ns19,shahbazyan-prb20} which, with further increase of $g/\gamma_{n}$, transforms into Rabi splitting, signaling the system's transition to strong coupling regime. From the Lorentzian model (\ref{pol-L-antenna2}), the complex frequencies of polaritonic states are $\omega_{\pm}=\omega_{n}-i(\gamma_{n}+\gamma_{0})/4\pm \sqrt{g^{2}-(\gamma_{n}-\gamma_{0})^{2}/16}$, implying that, for $\gamma_{0}/\gamma_{n}\ll 1$, strong coupling transition occurs at $g\gtrsim \gamma_{n}/4$. For smaller coupling, both polaritonic bands are centered at the same frequency $\omega_{n}$ whereas the ExIT minimum is due to energy transfer from the LSP to QE \cite{shahbazyan-prb20} within QE's narrow absorption spectral width $\gamma_{0}$.
%
\begin{figure}[tb]
\begin{center}
\includegraphics[width=0.99\columnwidth]{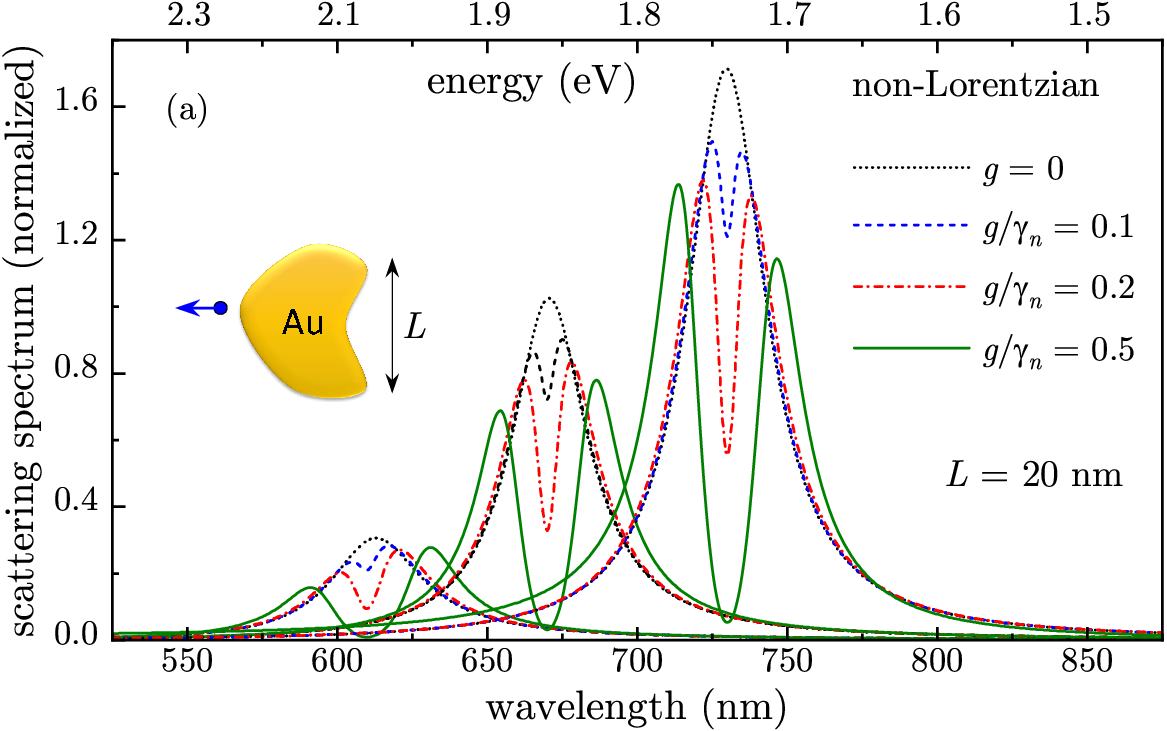}

\vspace{3mm}

\includegraphics[width=0.99\columnwidth]{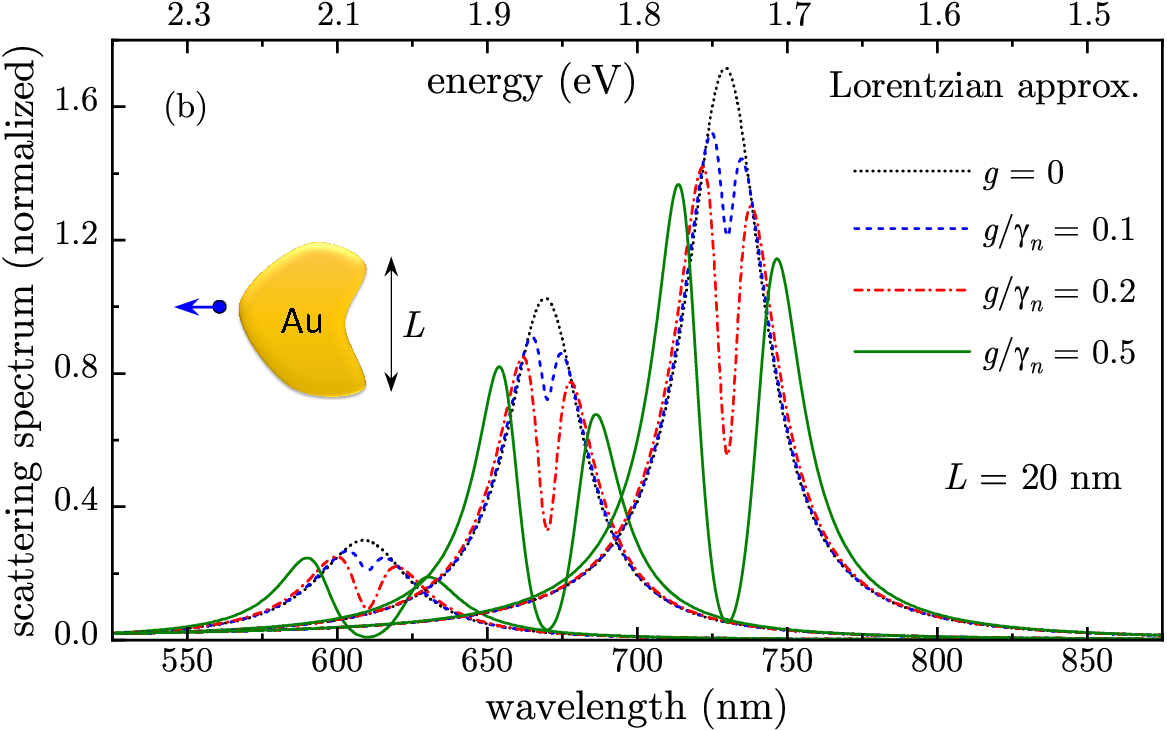}
\caption{\label{fig4} Normalized scattering cross-section $\sigma_{\rm scatt}/L^{2}$ calculated using non-Lorentzian model (a) and its Lorentzian approximation (b) for a QE near Au NP in water at LSP wavelengths 610 nm, 670 nm, and 730 nm. Inset: Schematics of a QE near Au NP of irregular shape.
 }
\end{center}
\vspace{-4mm}
\end{figure}
%

The striking difference between the spectra calculated using non-Lorentzian model and its Lorentzian approximation is the relative enhancement of lower energy polaritonic band clearly visible in Fig.~\ref{fig3}(a) for wavelengths below 700 nm, whereas in the Lorentzian case, similar to the CO model, the upper energy polaritonic band is enhanced in the entire spectral range [see Fig.~\ref{fig3}(b)]. This change of spectral asymmetry pattern persists at any value of coupling $g$ as the system transitions from the weak to strong coupling regime, indicating that this is a general non-Lorentzian effect. This effect is even more visible in scattering spectra, shown in Fig.~\ref{fig4} for the same set of system parameters, which exhibit a prominent but opposite asymmetry pattern in the wavelengths region below 700 nm for normalized scattering cross-section $\sigma_{\rm scatt}/L^{2}$ calculated using non-lorentzian model in Fig.~\ref{fig4}(a), and its Lorentzian approximation in Fig.~\ref{fig4}(b). These results are consistent with the recent experiment on tip-enhanced strong-coupling spectroscopy of a single QE  \cite{pelton-sci-adv19}.

Qualitatively, the enhancement of lower energy polaritonic band originates from the presence of frequency-dependent metal dielectric function $\varepsilon(\omega)$ in the numerator of effective polarizability  Eq.~(\ref{pol-antenna2}), whose real part behaves as $\varepsilon'(\omega)\propto \omega^{-2}$, resulting in a suppression of the upper energy polaritonic band. At the same time, the asymmetry reversal for wavelengths above 700 nm seen in Figs.~\ref{fig3}(a) and \ref{fig4}(a) can be traced to non-monotonic frequency dependence of the imaginary part of Au dielectric function  $\varepsilon''(\omega)$, as revealed by the LSP quality factor shown in Fig.~\ref{fig1}(b). Note that the Lorentzian polarizability (\ref{pol-L-antenna2}) has no dependence on $\varepsilon(\omega)$ and so the spectral weights of polaritonic bands are determined solely by the powers of $\omega$ in the expressions (\ref{cross-sections}) for extinction and scattering cross-sections, leading to enhancement of the upper energy polaritonic band in Figs.~\ref{fig3}(b) and \ref{fig4}(b).

\section{Conclusions}
\label{sec:conc}

In summary, we have developed non-Lorentzian model for quantum emitters (QE) resonantly coupled to localized surface plasmons (LSP) in metal-dielectric structures. Using the explicit form LSP Green function in the quasistatic limit, we derived non-Lorentzian extension of Maxwell-Bloch equations  describing LSP directly in terms of metal complex dielectric function rather than via Lorentzian resonances. For a single QE coupled to ta resonant LSP mode, we obtained an explicit expression for the system effective optical polarizability which, in the Lorentzian approximation, recovers the classical coupled oscillator (CO) model. We demonstrated that non-Lorentzian effects originating from the temporal dispersion of metal dielectric function affect significantly the optical spectra as the system transitions to strong coupling regime. Specifically, in the plasmonic frequency range, the main spectral weight is shifted towards the lower energy polaritonic band, consistent with the experiment.

In this paper, we have considered the role of non-Lorentzian effects in extinction and scattering spectra with numerical calculations performed for a single QE. In doing so, we were motivated by a recent experiment \cite{pelton-sci-adv19} on a single quantum dot in a plasmonic cavity which minimized other possible sources of the observed asymmetry pattern such as vibrons or interference effects. We expect that similar non-Lorentzian effects should be present in the emission spectra as well although, for a single QE, these are more difficult to observe in the absence of exciton-induced transparency minimum.

\acknowledgments
This work was supported in part by the National Science Foundation grants DMR-2000170, DMR-2301350, and NSF-PREM-2423854.



\begin{thebibliography}{99}

\bibitem{ebbesen-prl11} T. Schwartz, J. A. Hutchison, C. Genet, and T. W. Ebbesen,
\textit{Reversible Switching of Ultrastrong Light-Molecule Coupling},
Phys. Rev. Lett. \textbf{106}, 196405 (2011).

\bibitem{bachelot-nl13} A.-L. Baudrion, A. Perron, A. Veltri, A. Bouhelier, P.-M. Adam, and R. Bachelot,
\textit{Reversible Strong Coupling in Silver Nanoparticle Arrays Using Photochromic Molecules},
Nano Lett. \textbf{13}, 282–286 (2013).

\bibitem{zheng-nl16} L. Lin, M. Wang, X. Wei, X. Peng, C. Xie, and Y. Zheng, 
\textit{Photoswitchable Rabi Splitting in Hybrid Plasmon–Waveguide Modes},
Nano Lett. \textbf{16}, 7655–7663 (2016).

\bibitem{waks-nnano16} S. Sun, H. Kim, G. S. Solomon, and  E. Waks,
\textit{A quantum phase switch between a single solid-state spin and a photon}, 
Nat. Nanotechnol. \textbf{11}, 539–544 (2016).


\bibitem{senellart-nnano17} L. De Santis, C. Anton, B. Reznychenko, N. Somaschi, G. Coppola, J. Senellart, C. Gomez, A. Lemaitre, I. Sagnes, A. G. White, \textit{et al.,}
\textit{A solid-state single-photon filter},
Nat. Nanotech \textbf{12}, 663–667 (2017).
 
\bibitem{leggett-nl16} A. Tsargorodska, M. L. Cartron, C. Vasilev, G.Kodali, O. A. Mass, J. J. Baumberg, P. L. Dutton, C. N. Hunter, P.  T\"{o}rm\"{a}, and G. J. Leggett,
\textit{Strong Coupling of Localized Surface Plasmons to Excitons in Light-Harvesting Complexes}, 
Nano Lett. \textbf{16},  6850-6856 (2016).

\bibitem{shahbazyan-nl19} T. V. Shahbazyan, 
\textit{Exciton–Plasmon Energy Exchange Drives the Transition to a Strong Coupling Regime},
Nano Lett.  \textbf{19}, 3273–3279 (2019).


\bibitem{mortensen-rpp20}
C. Tserkezis, A. I. Fernandez-Dominguez, P. A. D. Goncalves, F. Todisco, J. D. Cox, K. Busch, N. Stenger, S. I. Bozhevolnyi, N. A. Mortensen, and C. Wolff,
\textit{On the applicability of quantum-optical concepts in strong-coupling nanophotonics},
Rep. Prog. Phys. \textbf{83}, 082401 (2020).

\bibitem{novotny-book} L. Novotny and B. Hecht, \textit{Principles of Nano-Optics} (CUP, New York, 2012).




\bibitem{forchel-nature04} J. P. Reithmaier, G. Sek, A. L\"{o}ffler, C. Hofmann, S. Kuhn, S. Reitzenstein, L. V. Keldysh, V. D. Kulakovskii, T. L. Reinecke, and A. Forchel,
\textit{Strong coupling in a single quantum dot–semiconductor microcavity system},
Nature \textbf{432}, 197 (2004).

\bibitem{khitrova-nphys06} G. Khitrova, H. M. Gibbs, M. Kira, S. W. Koch, and A. Scherer, 
\textit{Vacuum Rabi splitting in semiconductors},
Nature Phys. \textbf{2}, 81 (2006).

\bibitem{imamoglu-nature06}
K. Hennessy, A. Badolato, M. Winger, D. Gerace, M. Atat\"{u}re, S. Gulde, S. F\"{a}lt, E. L. Hu, and A. Imamoglu,
\textit{Quantum nature of a strongly coupled single quantum dot–cavity system},
Nature \textbf{445}, 896 (2006).


\bibitem{bellessa-prl04} J. Bellessa, C. Bonnand, J. C. Plenet, and J. Mugnier, 
\textit{Strong Coupling between Surface Plasmons and Excitons in an Organic Semiconductor},
Phys. Rev. Lett. \textbf{93}, 036404 (2004).

\bibitem{sugawara-prl06} Y. Sugawara, T. A. Kelf, J. J. Baumberg, M. E. Abdelsalam, and P. N. Bartlett, 
\textit{Strong Coupling between Localized Plasmons and Organic Excitons in Metal Nanovoids},
Phys. Rev. Lett. \textbf{97}, 266808 (2006).

\bibitem{wurtz-nl07} G. A. Wurtz, P. R. Evans, W. Hendren, R. Atkinson, W. Dickson, R. J. Pollard, A. V. Zayats, W. Harrison, and C. Bower,
\textit{Molecular Plasmonics with Tunable Exciton-Plasmon Coupling Strength in J-Aggregate Hybridized Au Nanorod Assemblies},
 Nano Lett. \textbf{7}, 1297 (2007).

\bibitem{fofang-nl08} N. T. Fofang, T.-H. Park, O. Neumann, N. A. Mirin, P. Nordlander, and N. J. Halas, 
\textit{Plexcitonic Nanoparticles: Plasmon-Exciton Coupling in Nanoshell-J-Aggregate Complexes},
Nano Lett. \textbf{8}, 3481 (2008).

\bibitem{bellessa-prb09} J. Bellessa, C. Symonds, K. Vynck, A. Lemaitre, A. Brioude, L. Beaur, J. C. Plenet, P. Viste, D. Felbacq, E. Cambril, and P. Valvin, 
\textit{Giant Rabi splitting between localized mixed plasmon-exciton states in a two-dimensional array of nanosize metallic disks in an organic semiconductor},
Phys. Rev. B \textbf{80}, 033303 (2009).



\bibitem{schlather-nl13}A. E. Schlather, N. Large, A. S. Urban, P. Nordlander, and N. J. Halas, 
\textit{Near-Field Mediated Plexcitonic Coupling and Giant Rabi Splitting in Individual Metallic Dimers},
Nano Lett. \textbf{13}, 3281 (2013).


\bibitem{lienau-acsnano14} W. Wang, P. Vasa, R. Pomraenke, R. Vogelgesang, A. De Sio, E. Sommer, M. Maiuri, C. Manzoni, G. Cerullo, and C. Lienau, 
\textit{Interplay between Strong Coupling and Radiative Damping of Excitons and Surface Plasmon Polaritons in Hybrid Nanostructures},
ACS Nano \textbf{8}, 1056 (2014).


\bibitem{shegai-prl15} G. Zengin, M. Wers\"{a}ll, S. Nilsson, T. J. Antosiewicz, M. K\"{a}ll, T. Shegai,
\textit{Realizing strong light-matter interactions between single nanoparticle plasmons and molecular excitons at ambient conditions},
Phys. Rev. Lett. \textbf{114}, 157401 (2015).



\bibitem{hakala-prl09} T. K. Hakala, J. J. Toppari, A. Kuzyk, M. Pettersson, H. Tikkanen, H. Kunttu, and P. Torma, 
\textit{Vacuum Rabi Splitting and Strong-Coupling Dynamics for Surface-Plasmon Polaritons and Rhodamine 6G Molecules},
Phys. Rev. Lett. \textbf{103}, 053602 (2009).

\bibitem{berrier-acsnano11} A. Berrier, R. Cools, C. Arnold, P. Offermans, M. Crego-Calama, S. H. Brongersma, and J. Gomez-Rivas, 
\textit{Active Control of the Strong Coupling Regime between Porphyrin Excitons and Surface Plasmon Polaritons},
ACS Nano \textbf{5}, 6226 (2011).

\bibitem{salomon-prl12} A. Salomon, R. J. Gordon, Y. Prior, T. Seideman, and M. Sukharev, 
\textit{Strong Coupling between Molecular Excited States and Surface Plasmon Modes of a Slit Array in a Thin Metal Film},
Phys. Rev. Lett. \textbf{109}, 073002 (2012).


\bibitem{luca-apl14}A. De Luca, R. Dhama, A. R. Rashed, C. Coutant, S. Ravaine, P. Barois, M. Infusino, and G. Strangi,
\textit{Double strong exciton-plasmon coupling in gold nanoshells infiltrated with fluorophores}, 
Appl. Phys. Lett. \textbf{104}, 103103 (2014).

\bibitem{noginov-oe16} V. N. Peters, T. U. Tumkur, Jing Ma, N. A. Kotov, and M. A. Noginov,
\textit{Strong coupling of localized surface plasmons and ensembles of dye molecules}, 
Opt. Express \textbf{24}, 25653 (2016).

\bibitem{vasa-prl08} P. Vasa, R. Pomraenke, S. Schwieger, Y. I. Mazur, V. Kunets, P. Srinivasan, E. Johnson, J. E. Kihm, D. S. Kim, E. Runge, G. Salamo, and C. Lienau, 
\textit{Coherent Exciton-Surface-Plasmon-Polariton Interaction in Hybrid Metal-Semiconductor Nanostructures},
Phys. Rev. Lett. \textbf{101}, 116801 (2008).

\bibitem{gomez-nl10} D. E. Gomez, K. C. Vernon, P. Mulvaney, and T. J. Davis, 
\textit{Surface Plasmon Mediated Strong Exciton-Photon Coupling in Semiconductor Nanocrystals},
Nano Lett. \textbf{10}, 274 (2010).

\bibitem{gomez-jpcb13} D. E. Gomez, S. S. Lo, T. J. Davis, and G. V. Hartland, 
\textit{Picosecond Kinetics of Strongly Coupled Excitons and Surface Plasmon Polaritons},
J. Phys. Chem. B \textbf{117}, 4340 (2013).


\bibitem{manjavacas-nl11} A. Manjavacas, F. J. Garcia de Abajo, and P. Nordlander, 
\textit{Quantum Plexcitonics: Strongly Interacting Plasmons and Excitons},
Nano Lett. \textbf{11}, 2318 (2011).





\bibitem{hecht-sci-adv19} H. Gross, J. M. Hamm, T. Tufarelli, O. Hess, and  B. Hecht,
\textit{Near-field strong coupling of single quantum dots},
Sci. Adv. \textbf{4}, eaar4906 (2018).

\bibitem{pelton-sci-adv19} K.-D. Park, M. A. May, H. Leng, J. Wang, J. A. Kropp, T. Gougousi, M. Pelton, and M. B. Raschke,
\textit{Tip-enhanced strong coupling spectroscopy, imaging, and control of a single quantum emitter},
Sci. Adv. \textbf{5}, eaav5931 (2019).

\bibitem{baumberg-natmat2019} J. J. Baumberg, J. Aizpurua, M. H. Mikkelsen, and D. R. Smith, 
\textit{Extreme nanophotonics from ultrathin metallic gaps},
Nat. Mater.  \textbf{18},  668–678 (2019).



\bibitem{pelton-oe10} X. Wu, S. K. Gray, and M. Pelton,
\textit{Quantum-dot-induced transparency in a nanoscale plasmonic resonator},
Optics Express \textbf{18},  23633-23645 (2010).

\bibitem{pelton-nc18} H. Leng, B. Szychowski, M.-C. Daniel, and   M. Pelton,
\textit{Strong coupling and induced transparency at room temperature with single quantum dots and gap plasmons},
Nat. Commun. \textbf{9}, 4012 (2018).

\bibitem{pelton-ns19} M. Pelton, S. D. Storm,  and  H.  Leng,
\textit{Strong coupling of emitters to single plasmonic nanoparticles: Exciton-induced transparency and Rabi splitting},
Nanoscale \textbf{11}, 14540-14552 ( 2019).





\bibitem{savvidis-aom13} N. Christogiannis, N. Somaschi, P. Michetti, D. M. Coles, P. G. Savvidis, P. G. Lagoudakis, and D. G. Lidzey, 
\textit{Characterizing the electroluminescence emission from a strongly coupled organic semiconductor microcavity LED}, 
Adv. Opt. Mater. \textbf{1}, 503 (2013).

\bibitem{ebbesen-fd15} J. George, S. Wang, T. Chervy, A. Canaguier-Durand, G. Schaeffer, J.-M. Lehn, J. A. Hutchison, C. Genet, and T. W. Ebbesen, 
\textit{Ultra-strong coupling of molecular materials: spectroscopy and dynamics},
Faraday Discuss. \textbf{178}, 281–294 (2015).

\bibitem{ebbesen-nc15} A. Shalabney, J. George, J. Hutchison, G. Pupillo, C. Genet and T. W. Ebbesen, 
\textit{Coherent coupling of molecular resonators with a microcavity mode}, 
Nature Comm. \textbf{6}, 5981 (2015).


\bibitem{shegai-nl17} M. Wers\"{a}ll, J. Cuadra, T. J. Antosiewicz, S. Balci, and T. Shegai,
\textit{Observation of mode splitting in photoluminescence of individual plasmonic nanoparticles strongly coupled to molecular excitons},
Nano Lett. \textbf{17}, 551-558 (2017).


\bibitem{shegai-acsphot19} M. Wers\"{a}ll, B. Munkhbat, D. G. Baranov, F. Herrera, J. Cao, T. J. Antosiewicz, and T. Shegai,
\textit{Correlative Dark-Field and Photoluminescence Spectroscopy of Individual Plasmon–Molecule Hybrid Nanostructures in a Strong Coupling Regime},
ACS Photonics  \textbf{6}, 2570–2576 (2019).


\bibitem{zhang-nl17}
D. Zheng, S. Zhang, Q. Deng, M. Kang, P. Nordlander, H. Xu,
\textit{Manipulating Coherent Plasmon-Exciton Interaction in a Single Silver Nanorod on Monolayer WSe$_2$},
Nano Lett. \textbf{17}, 3809  (2017).


\bibitem{xu-nl17} J. Wen, H. Wang, W. Wang, Z. Deng, C. Zhuang, Y. Zhang, F. Liu, J. She, J. Chen, H. Chen, \textit{et al.,}
\textit{Room-Temperature Strong Light–Matter Interaction with Active Control in Single Plasmonic Nanorod Coupled with Two-Dimensional Atomic Crystals},
Nano Lett. \textbf{17},  4689 (2017).


\bibitem{garsia-vidal-njp15} J. del Pino, J. Feist, and F. J. Garcia-Vidal, 
\textit{Quantum theory of collective strong coupling of molecular vibrations with a microcavity mode},
New J. Phys. \textbf{17}, 053040 (2015).

\bibitem{aizpurua-optica18}T. Newman and J. Aizpurua,
\textit{Origin of the asymmetric light emission from molecular exciton–polaritons},
Optica \textbf{5}, 1247 (2018).




\bibitem{ding-prl17}
S.-J. Ding, X. Li, F. Nan, Y.-T. Zhong, L. Zhou, X. Xiao, Q.-Q. Wang, and Z. Zhang
\textit{Strongly Asymmetric Spectroscopy in Plasmon-Exciton Hybrid Systems due to Interference-Induced Energy Repartitioning},
Phys. Rev. Lett. \textbf{119}, 177401 (2017).

\bibitem{shahbazyan-nanophot21}T. V. Shahbazyan,
\textit{Transition to strong coupling regime in hybrid plasmonic systems: Exciton-induced transparency and Fano interference},
Nanophotonics \textbf{10}, 3735 (2021).


\bibitem{xu-acsphot21} F. Shen, Z. Chen, L. Tao, B. Sun, X. Xu, J. Zheng, and J. Xu,
\textit{Investigation on the Fano-Type Asymmetry in Atomic Semiconductor Coupled to the Plasmonic Lattice},
ACS Photon.  \textbf{8}, 212 (2021).


\bibitem{shahbazyan-jcp22} Z. Scott, S. Muhammad, and T. V. Shahbazyan,
\textit{Plasmon-induced coherence, exciton-induced transparency, and Fano interference for hybrid plasmonic systems in strong coupling regime},
J. Chem. Phys. \textbf{156}, 194702 (2022).

\bibitem{shahbazyan-prb22}T. V. Shahbazyan,
\textit{Non-Markovian effects for hybrid plasmonic systems in the strong coupling regime},
Phys. Rev. B \textbf{105}, 245411 (2022).

\bibitem{welsch-pra98}
H. T. Dung, L. Kn\"{o}ll, and D.-G. Welsch, 
\textit{Three-dimensional quantization of the electromagnetic field in dispersive and absorbing inhomogeneous dielectrics}, 
Phys. Rev. A \textbf{57}, 3931 (1998).


\bibitem{philbin-njp10}
T. G. Philbin,
\textit{Canonical quantization of macroscopic electromagnetism},
New J. Phys. \textbf{12}  123008 (2010).

\bibitem{shahbazyan-prb21} T. V. Shahbazyan, 
\textit{Interacting quantum plasmons in metal-dielectric structures},
Phys. Rev. B \textbf{103}, 045421 (2021).


\bibitem{shahbazyan-pra23} T. V. Shahbazyan, 
\textit{Universal optical polarizability for plasmonic nanostructures},
Phys. Rev. A \textbf{107}, L061503 (2023).






\bibitem{stockman-review} M. I. Stockman, 
\textit{Nanoplasmonics: From Present into Future}, 
in \textit{Plasmonics: Theory and Applications}, edited by T. V. Shahbazyan and M. I. Stockman (Springer, New York, 2013).






\bibitem{shahbazyan-prl16} T. V. Shahbazyan, 
\textit{Local density of states for nanoplasmonics},
Phys. Rev. Lett. \textbf{117}, 207401 (2016).

\bibitem{shahbazyan-prb18} T. V. Shahbazyan, 
\textit{Spontaneous decay of a quantum emitter near a plasmonic nanostructure},
Phys. Rev. B \textbf{98}, 115401 (2018).

\bibitem{shahbazyan-acsphot17} T. V. Shahbazyan, 
\textit{Mode volume, energy transfer, and spaser threshold in plasmonic systems with gain},
ACS Photon. \textbf{4}, 1003 (2017).

\bibitem{shahbazyan-prb20} T. V. Shahbazyan, 
\textit{Exciton-induced transparency in hybrid plasmonic systems},
Phys. Rev. B \textbf{102}, 205409 (2020).


\end{thebibliography}
\end{document}